# Non-conventional porous 12CaO·7Al2O3 Transparent Oxide Semiconductor: A Journey from foams to cubes

Deepak Dwivedi*, Vaibhav Bhavsar, Aditi Thanki

Department of Materials Science & Engineering, Indian Institute of Technology Gandhinagar, 382424, Chandkheda, Ahmedabad, India

*Corresponding author E-mail address: deepakdwivedi007@yahoo.com

**Abstract:**

In the current study easy method to synthesize C12A7 particles with cubic morphology was investigated for the application of Transparent Conducting Oxide layer in thinfilm solar cell. By self-combustion method with proper amount of urea addition, cubic and foamy structures of C12A7 were achieved at 950 °C and at 510 °C respectively. Pure C12A7 formation confirmed by P-XRD and growth mechanism of the cubes was also understood by SEM and FT-IR analysis. Further the band gap was measured 3.07 eV in case of 510 °C and 3.32 eV in case of 950 °C synthesized particles with the help of UV-Vis analysis. During PL single sharp peaks at 399 nm noticed which indicates less defective C12A7 at 950 °C, confirmed by XRD having more crystallinity and less strain. No trap state or radiative recombination observed in case of C12A7 synthesized at 950 °C sample. Conductivity from conductometry titration method was found to be 9.223 mS at 26.3 °C and 747.33 mS at 26.4 °C for 510 °C and 950 °C synthesized samples respectively.

## 1. Introduction

Transparent conductive electrodes (TCEs), which transmit light in the visible spectral range and conduct electricity simultaneously, are of increasing importance for information (i.e., displays, touch panels) and energy (i.e., photovoltaics, architectural and window glass) technologies[1], [2]. It is believed that combination of high optical transparency and high electronic conduction is contradictory, since optical transparent material is an insulator with band gaps larger than 3.3 eV and this large gap makes carrier doping difficult. This combination is achieved in several commonly used oxides - $In_2O_3$, $SnO_2$, $TiO_2$, ZnO and CdO. To become a transparent conducting oxide (TCO), these hosts must be degenerately doped to displace the Fermi level up into the conduction band. Currently, Sn-doped $In_2O_3$ (ITO), F-doped $SnO_2$ (FTO) and Al-doped ZnO (AZO) are the most widely used TCOs because of their low resistivity (on the order of $10^{-4}$ Ω cm), high optical transmittance (80-90%) and ease of fabrication[3]–[7]. Recently, the metallic and superconducting states has been observed in a typical refractory oxide $12CaO \cdot 7Al_2O_3$ (C12A7) with built-in porous structure[8]. In contrast to conventional TCOs, where there is trade-off between optical transparency and electrical conductivity, porous materials offer possibility to combine 100% optical transparency with high electrical conductivity. In addition, C12A7 is made from the abundant and typical insulator CaO and $Al_2O_3$. C12A7 has cubic lattice with space group of $I\bar{4}3d$ and lattice constant of 1.199nm[8]. The unit cell composed of positively-charged lattice framework $[Ca_{24}Al_{28}O_{64}]^{4+}$ and two extra-framework $O^{2-}$ ions called 'free oxygen ions'. The framework has twelve subnanometer sized cages having a free space of 0.4 nm in diameter three-dimensionally connected with each other through a monomolecular thick cage wall. Two out of twelve cages are occupied by the free oxygen ions while remaining ten cages are empty, resulting in the formation of cage conduction band (CCB). The replacement of loosely bound free $O^{2-}$ ion with other monovalent anions, such as hydroxyl ions ($OH^-$) or halogen anions $X^-$ ($F^-$ and $Cl^-$), stabilizes the structure through charge delocalization[9]–[11]. Although the crystal structure looks rather complex, the synthesis of C12A7 is quite easy. C12A7 can be synthesized by conventional solid state reactions in air using $CaCO_3$ and $Al(OH)_3$ or $Al_2O_3$ by heating at temperature > 1000 °C[12]. In this work, we have synthesized C12A7 powder by solution combustion route. The synthesized powder was characterized by X-ray diffraction (XRD), scanning electron microscopy, etc.

## 2. Experimental Procedure:

### 2.1 Synthesis:

C12A7 particles were synthesized by self-propogating combustion synthesis process[12]. Calcium nitrate tetrahydrate ($Ca(NO_3)_2 \cdot 4H_2O$, EMSURE ACS, 99%, MERCK) was used as Ca source and Aluminum nitrate nonahydrate ($Al(NO_3)_2 \cdot 9H_2O$, EMSURE, MERCK) was used as Al source. Ultra-high purity water (18.2 MΩ-cm) was obtained from a Millipore Direct-Q water purification system. Urea (GR for analysis ACS, MERCK) was used as a fuel for combustion during the synthesis of C12A7 particles. For synthesis process first 0.4 M stock solutions of $Ca(NO_3)_2 \cdot 4H_2O$ and $Al(NO_3)_2 \cdot 9H_2O$ was prepared by addition of 11.81 gm and 18.76 gm $Ca(NO_3)_2 \cdot 4H_2O$ and $Al(NO_3)_2 \cdot 9H_2O$ in 125 ml water respectively. 1.5 grams of urea was then added to both the stock solutions. In order to get the cation solution (250 ml) now both the stock solutions are mixed in a Pyrex™ beaker and stirred with magnetic stirrer for an hour at room temperature. Final solution was then placed in muffle furnace and first soaked at a temperature of 510º ± 10 ºC for 30 minutes to allow the completion of combustion process. $1^{st}$ sample product was taken out from the furnace and allowed to cool at room temperature. This results in the foamy product of the desired calcium aluminate phase and then lightly grounded in the mortar to get fine powder. $2^{nd}$ sample was kept inside the muffle furnace up to 910ºC ± 10ºC for completion of the calcination process and then furnace cooled. This was resulted in the crystalline product which was also grounded in mortar to get the fine powder product.

**2.2 Microstructural Characterization:**

The detailed microstructural characterization of these as prepared Calcium aluminate particles were carried out using experimental techniques, such as, Powder X-ray Diffractometer (P-XRD; *Model: D8 Discover, Supplier: Bruker Corporation*) and Field Emission Scanning Electron Microscopy (FE-SEM; *Model: JSM 7600F, Supplier: JEOL Ltd.*). Whereas P-XRD is used to know the crystallinity, phase and crystallite size, etc. The morphology of these particles is determined using FE-SEM. The P-XRD measurements were carried out in the 2θ range of 10 - 90° (step size of 0.02°) by Cu $K_\alpha$ ($\lambda$ = 0.15418 nm) radiation. For FE-SEM analysis, first the powder particles were dispersed in water and then coated with Pt before introducing them into the FE-SEM chamber. Accelerating voltage of 5 kV was used to capture these images.

**2.3 FT-IR:**

A Fourier transformation infrared spectrometer (*Model: Nicolet iS10, Supplier: Thermo Scientific*) was used to investigate the chemical bonds formed in both foamy and crystalline product of calcium aluminates. Proper sample preparation method was followed and characterization was done very preciously.

**2.4 UV-Vis:**

Optical properties (absorbance and band gap) of synthesized calcium aluminate particles were investigated by UV-Spectrometer (*Model: Specord @ 210 plus analitic, Supplier: jena, Germeny*) in the wavelength of 200-800 nm. Initially the baseline was corrected with milli-Q ultra-pure water in silica quartz of 10 mm.

**2.5 Photoluminescence:**

Photoluminescence of both foamy and crystalline calcium aluminate was analysed by Fluorescence Spectroscopy (*Model: Flurolog JOBIN YVON, Supplier: Horiba Scientific*).

**2.6 Conductivity:**

Conductivity of the product was measured by conductometry titration method using conductivity meter (*Model: Conductivity meter 306, Supplier: SYSTRONICS Ltd.*)

**3. Results and Discussion:**

**3.1 XRD analysis:**

Figure 1 (a) shows the XRD graph for particles synthesized at 510 °C. It showed strong diffraction peaks at 2θ = 14.48°, 17.18°, 18.66°, 22.54°, 25.87°, 27.13°, 29.20°, 34.08°, 35.55°, 37.71°, 41.40°, 43.34°, 46.58°, 49.04°, 51.13°, 61.89°, 71.77 and 74.68° corresponding to (002), (110), (111), (112), (020), (202), (004), (023), (221), (311), (223), (313), (400), (402), (134), (334), (046) and (602) planes of $Ca(NO_3)_2·2H_2O$. This is in good agreement with the values of standard card (PDF 00-027-0087). Comparatively weak diffraction peaks at 2θ = 20.39°, 27.85°, 40.56°, 47.38°, 54.05° and 58.68° observed corresponding to (020), (111), (201), (122), (212) and (320) planes of $Al(OH)_3$. This is in good agreement with the values of standard card (PDF 00-020-0011) in the samples of particles synthesized at 510°C. This indicates that the complete evaporation of precursors used for the synthesis is not done at this temperature.

Figure 1 (b) shows the XRD graph for particles synthesized at 950 °C. It showed strong diffraction peaks at 2θ = 33.40°, 18.12°, 36.69°, 41.20°, 46.66°, 67.14°, 70.17°, 54.05°, 52.87°, 72.22°, 87.68°, 60.80° and 75.16° corresponding to (420), (211), (422), (521), (640), (611), (831), (840), (710), (444), (842), (1040), (732) and (930) planes of C12A7. This is in

good agreement with the values of standard card (PDF 00-009-0413). Comparatively weak diffraction peaks at 2θ = 42.36°, 47.68° and 59.34° observed corresponding to (711), (800) and (844) planes of C3A. This is in good agreement with the values of standard card (PDF 00-038-1429) in the samples of particles synthesized at 950°C. This indicates the complete evaporation of the precursors used for the synthesis and the presence of only dry CaO and AlO intermetalics.

The crystallite size and micro-strain were calculated using Williamson-Hall formula[13] and showed in table 1.

$$\beta \cos\theta = \frac{k\lambda}{t} + \eta \sin\theta \quad \ldots\ldots\ldots\ldots\ldots\ldots\ldots\ldots\ldots\ldots\ldots\ldots\ldots\ldots\ldots\ldots\ldots\ldots\ldots\ldots\ldots\ldots\ldots\ldots\ldots\ldots\ldots\ldots\ldots\ldots\ldots\ldots\ldots\ldots\ldots\ldots \quad (1)$$

Where λ is the wavelength of Cu-Kα X rays (λ = 0.15418 nm), $\beta$ is FWHM of diffraction peaks and θ is diffraction angle. The plot of $\beta \cos\theta$ versus $\sin\theta$ gives a straight line as shown in figure (2) with slope equal to $\eta$ and the intercept along y-axis as $k\lambda/t$, from which micro-strain and crystallite size can be measured.

### 3.2 FESEM analysis:

FESEM is capable to give exact information about the length scale parameter of microstructure like grain size, sub grain size, particle size and morphology of the as synthesized particles. SEM images of C12A7 particles synthesized at 510 °C and 950 °C are shown in figure (3) at different magnifications. Foamy feathery structure was observed in the case of 510 °C sample and very sharp edged cubes formed in the case of 950 °C sample.

For further investigation of the surfaces of C12A7 powder was scanned at higher magnification and shown in figure (4). In this two types of surfaces on cubes were observed which were flat and wavy (Fig. (4) a'-d') whereas foamy particle surface was observed same (Fig (4) a-d).

It was observed from the figure (4) that some of the cubes are very sharp with a clear edge formation whereas some of the cubes are bent and also contained other particles on the surfaces. Attachment of the particles on the surface could be due to the presence of high surface energy facets in the cubes. It was also noticed that none of the cube was hollow in nature but at 510°C all the particles were hollow and foamy with significant porosity. In the high temperature regime, the precursor atoms have enough energy to get mobile and react and diffuse properly to form a building block of the morphology.

In order to find out the mechanism of cube formation at 950°C, we took the support of FTIR data which explained the presence of $OH^-$ ions and H-O-H stretching in the particles synthesized at 550 °C whereas this was completely missing at 950 °C. This could be responsible to reduce the internal porosity in structure. At 950°C we observed the presence of $CN^-$ stretching whereas at 510 °C significant amount of $CH_2$ stretching (1349) was observed along with the NH stretching. Therefore we propose that for the formation of cube presence of $CN^-$ ions is critical and removal of $OH^-$ was responsible for complete crystallinity of particles at 950 °C.

Following would be the possible growth mechanism which could be graphically represented as in figure 5:

(a) At 950 °C, removal of $OH^-$ ion, NH and $CH_2$ helps to generate open volume through ejection and helps CN to come in structure as more evident peak was observed in FTIR.

(b) CN acted as seed along with the C=S and C-H stretching and high temperature supported ions to get mobile and diffuse with each other to form a building block.

(c) Due to the difference in the Ksp of the ions, diffusion path length and diffusion time was different at the different region due to the varying concentration of ions which lead to the formation of wavy surface of the cubes along with the attachment of other particles.

(d) This statement was further justified by FESEM image as the appearance of wavy surface in cube is due to the attachment of various ions on the surface after the completion of cube formation. This could be done either due to the – (1) high concentration of ions in solution which were left un reacted and at the end sticked with the surface of cubes and generated functionalized surface or (2) rough surface generation due to the high surface energy facets formation. Extensive study needs to make this confirm.

**3.3 UV-Vis analysis:**

Figure 6 (a), (b) shows the UV-Vis spectra for C12A7 samples synthesized at 510 °C and 950 °C. An excitonic absorption peak located at around 372 nm in case of 950 °C sample but there is no such peak in case of 510 °C sample. Indication of sharp peak in 950°C indicates that the material is crystalline in nature which has been also proved by earlier PXRD and FESEM results whereas at 510°C, foamy particles were observed with poor crystallinity which got further affirmed with UV-Vis analysis (Figure 5 (a)). Appearance of sharp excitonic peak also reveals about the band to band transition and at the same time capable to comment about the particle size in a qualitative way which could be further cross checked by XRD data. Crystallite size in 950°C was 96 nm with strain of 0.00153 whereas crystallite size of particles synthesized at 510°C had crystallite size 90 nm with strain of 0.00164. Released strain in particles synthesized at 950°C might be responsible for the smooth transition of electron from valence band to conduction band. At the same time, lesser strain of particles at 950°C also comments qualitatively about dislocation density or defects. In both the cases, positive strains were observed which is nothing but due to the tensile strain and it is an established fact that tensile strain sustains in a material due to the positive edge dislocation.

At higher temperature, atoms got enough energy to redefine their position inside the crystal structure which could lead to the reduction of strain. This study could also be useful for the material chemist because at 510 °C, large strain may support foreign atoms in the form of doping because of large open spaces available in the crystal structure. Extensive study is needed to establish this fact. Further band gap was measured 3.07eV in case of 510 °C sample and 3.32eV in case of 950 °C sample.

**3.4 FT-IR analysis:**

Figure 7 (a), (b) shows the FT-IR graphs for C12A7 samples analysed at 510 °C and 950 °C respectively. The band between 3200cm$^{-1}$ to 3650cm$^{-1}$ was attributed to the OH- vibration of the water molecules. The peaks observed at around 1500cm$^{-1}$ to 1600 cm$^{-1}$ are assigned to the vibration of the H-O-H bond. The band observed at 1000 cm$^{-1}$ to 1010 cm$^{-1}$ assigned as band due to sulphate group but Wang et al assigned band at 1020 cm$^{-1}$ as C-OH band. The peaks observed at 1500cm$^{-1}$ to 1600 cm$^{-1}$ are assigned to the vibration of the H-O-H bond Band at 1418 cm$^{-1}$ and 1551 cm$^{-1}$ can be assigned as the band due to the carboxylate group. 1480 cm$^{-1}$ and 1650 cm$^{-1}$ may assigned as C-N stretching and N-H banding. This also confirms the conclusion from XRD results that complete evaporation of precursors used for synthesis only take place at 950 °C and presence of only dry CaO and AlO intermetallics.

**3.5 PL analysis:**

In order to acquire a sound knowledge about the surface characteristic of C12A7 particles, PL emission spectrum of C12A7 nanoparticles synthesized at 950 °C. Since it is well established that temperature affects surface characteristics for getting information about the various states responsible for PL emission, the experimentally obtained spectrums are subjected to the multipeak Gaussian fitting (Figure 8 (b)). PL spectrums contain peaks between 368 nm to 502 nm as shown in figure 8 (a). These peaks were confirmed by us through repetition of experiments and we concluded that these peaks are not the artifact of instrument. After Gaussian fitting, we noticed single sharp peaks at 399 nm which indicates less defective C12A7 at 950 °C. This is also confirmed by XRD that C12A7 synthesized at 950 °C has more crystallinity and less strain and no trap state or radiative recombination observed in case of C12A7 synthesized at 950 °C sample.

### 3.6 Conductivity measurement:

Conductivity from conductometry titration method was found to be 9.223 mS at 26.3 °C and for 510 °C and 747.33 mS at 26.4 °C for 950 °C synthesized samples respectively.

### 4. Conclusion:

An attempt was made to explore a new transparent conductive oxide material synthesis and authors were successful to synthesize C12A7 with significant conductivity as obtained by conductometry titration. SEM results indicated the cube formation in 950°C whereas foamy particles were synthesized at 510°C. An important conclusion came out from the study that due to the lesser strain in particles synthesized at 950°C than the particles synthesized at 510 °C , it shown a band to band smooth transition which affirmed by UV-Vis analysis with a band gap of 3.32 eV. A significant approach was made to find out the best possible mechanism of cube formation. It was a challenge to bring high conductivity in material like C12A7 and this paper presented a method to generate good conductivity through morphology changing of particles. Vast and extensive study is open now to explore the role of cubes in conductivity enhancement of C12A7.

# Figure and Table Captions List

**FIGURE 1 :** X-ray diffractograms of C12A7 particles synthesized at (a) 510 °C and (b) 950 °C temperature.

**FIGURE 2:** Williumson Hall plot for C12A7 particles synthesized at (a) 510 °C and (b) 950 °C temperature.

**FIGURE 3:** Morphology of C12A7 particles synthesized at (a), (b) 510 °C and (c), (d) 950 °C temperature.

**FIGURE 4:** Surface morphology of C12A7 particles synthesized at (a), (b), (c), (d) 510 °C and (a')(b'), (c'),(d') 950 °C temperature.

**FIGURE 5:** Growth mechanism of C12A7 particles.

**FIGURE 6:** UV spectra of C12A7 particles synthesized at (a) 510 °C and (b) 950 °C temperature.

**FIGURE 7:** FT-IR spectra of C12A7 particles synthesized at (a) 510 °C and (b) 950 °C temperature.

**FIGURE 8:** PL spectrum and (b) multipeak Gaussian fitting of C12A7 particles synthesized at 950 °C.

**TABLE 1:** Crystallite size and micro-strain of C12A7 particles.

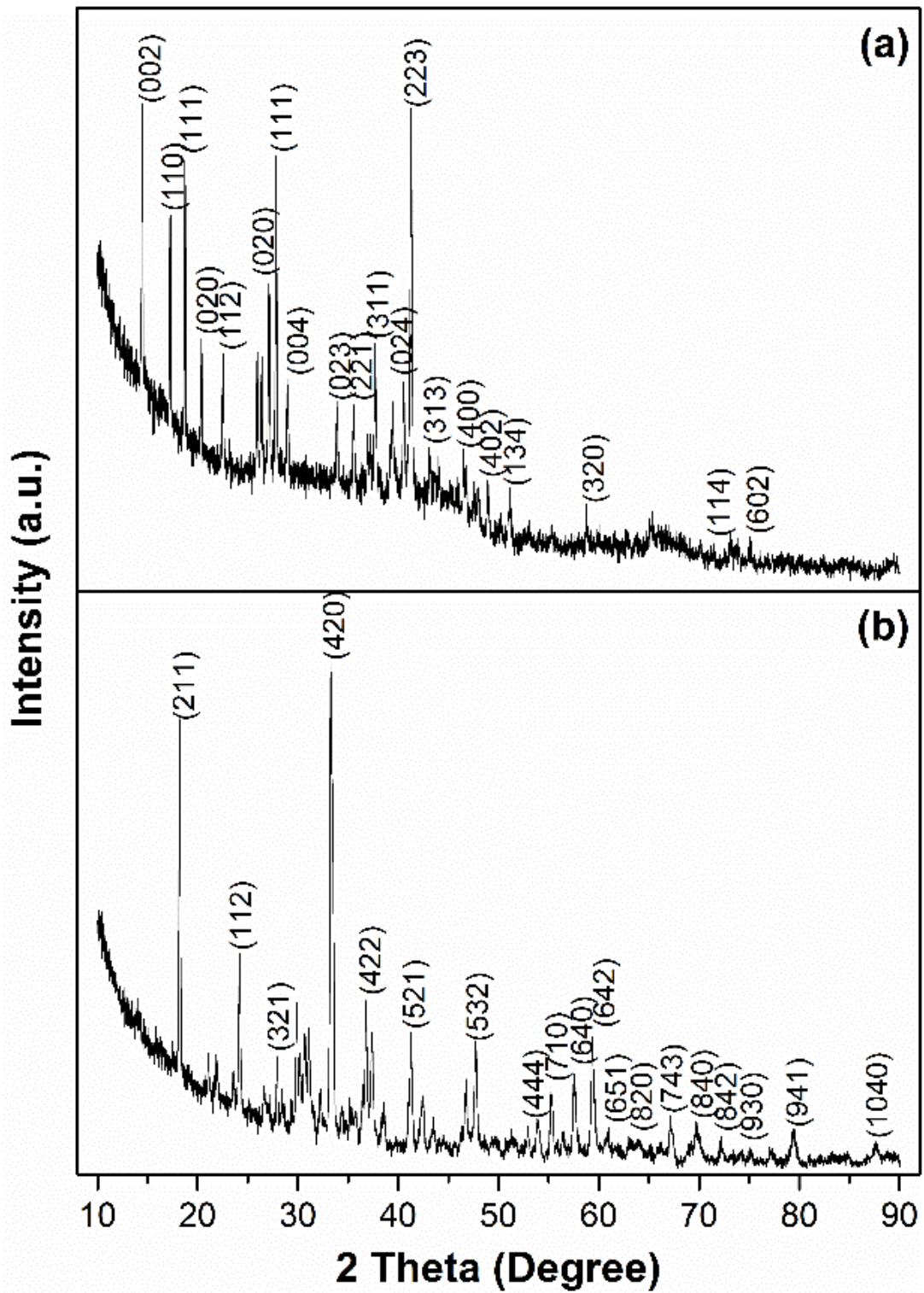

**FIGURE 1 :** X-ray diffractograms of C12A7 particles synthesized at (a) 510 °C and (b) 950 °C temperature.

**FIGURE 2:** Williumson Hall plot for C12A7 particles synthesized at (a) 510 °C and (b) 950 °C temperature.

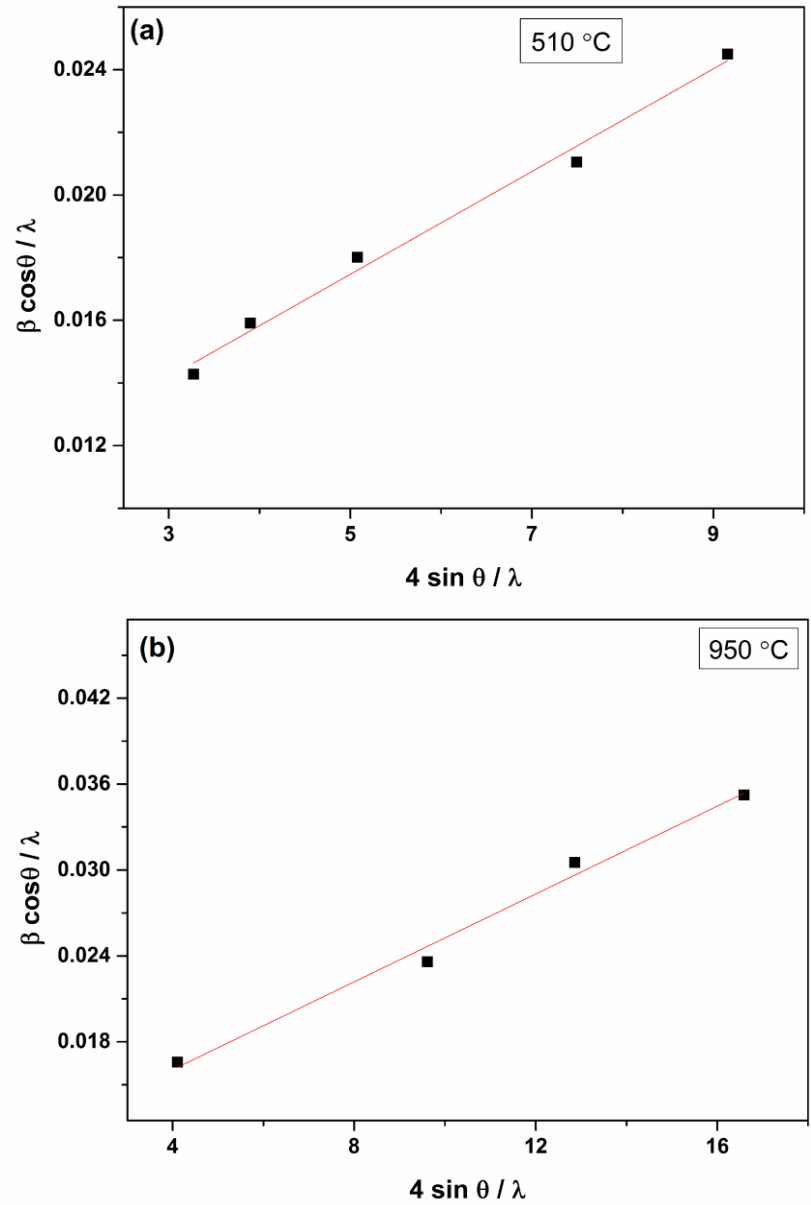

**FIGURE 3:** Morphology of C12A7 particles synthesized at (a), (b) 510 °C and (c), (d) 950 °C temperature.

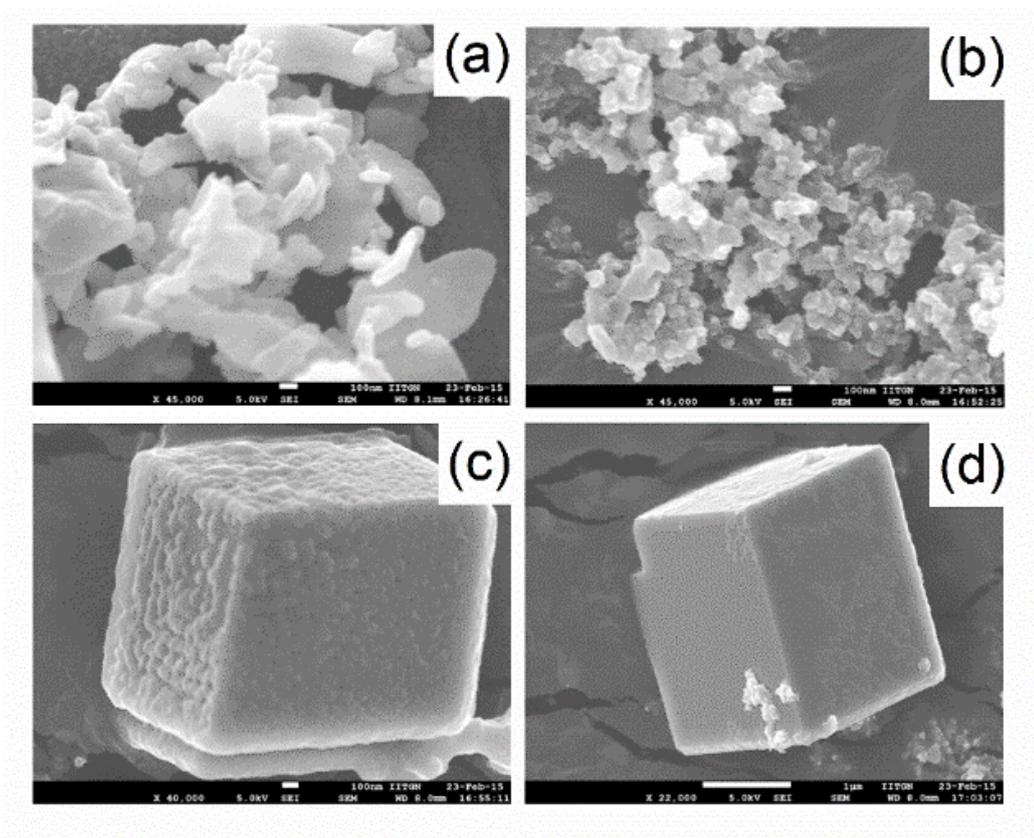

**FIGURE 4:** Surface morphology of C12A7 particles synthesized at (a), (b), (c), (d) 510 °C and (a')(b'), (c'),(d') 950 °C temperature.

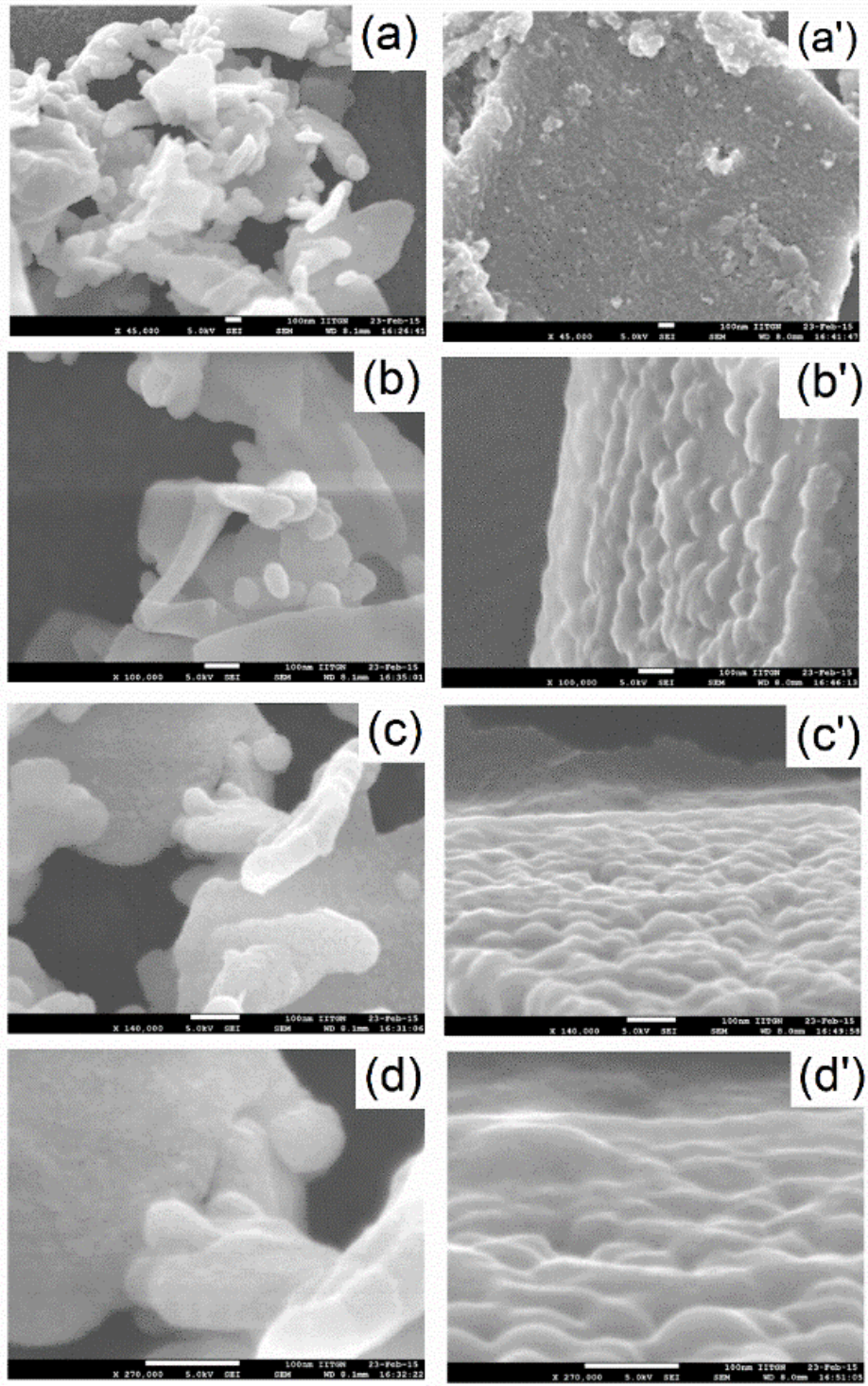

**FIGURE 5:** Growth mechanism of C12A7 particles.

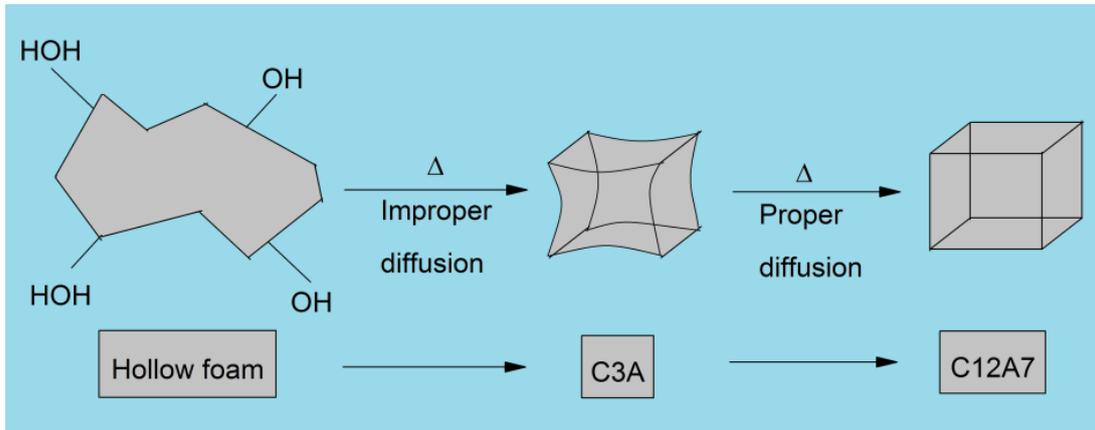

**FIGURE 6:** UV spectra of C12A7 particles synthesized at (a) 510 °C and (b) 950 °C temperature.

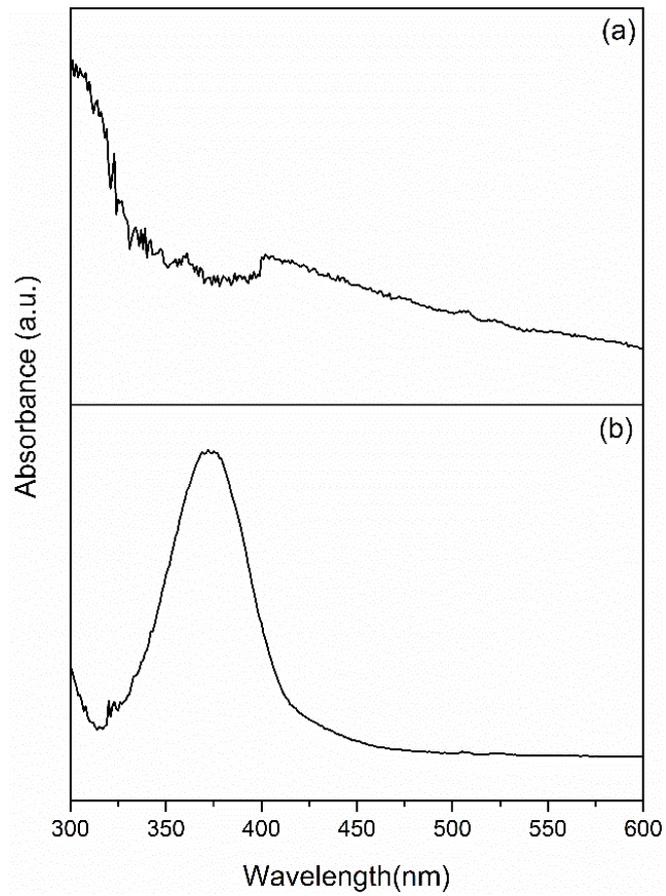

**FIGURE 7:** FT-IR spectra of C12A7 particles synthesized at (a) 510 °C and (b) 950 °C temperature.

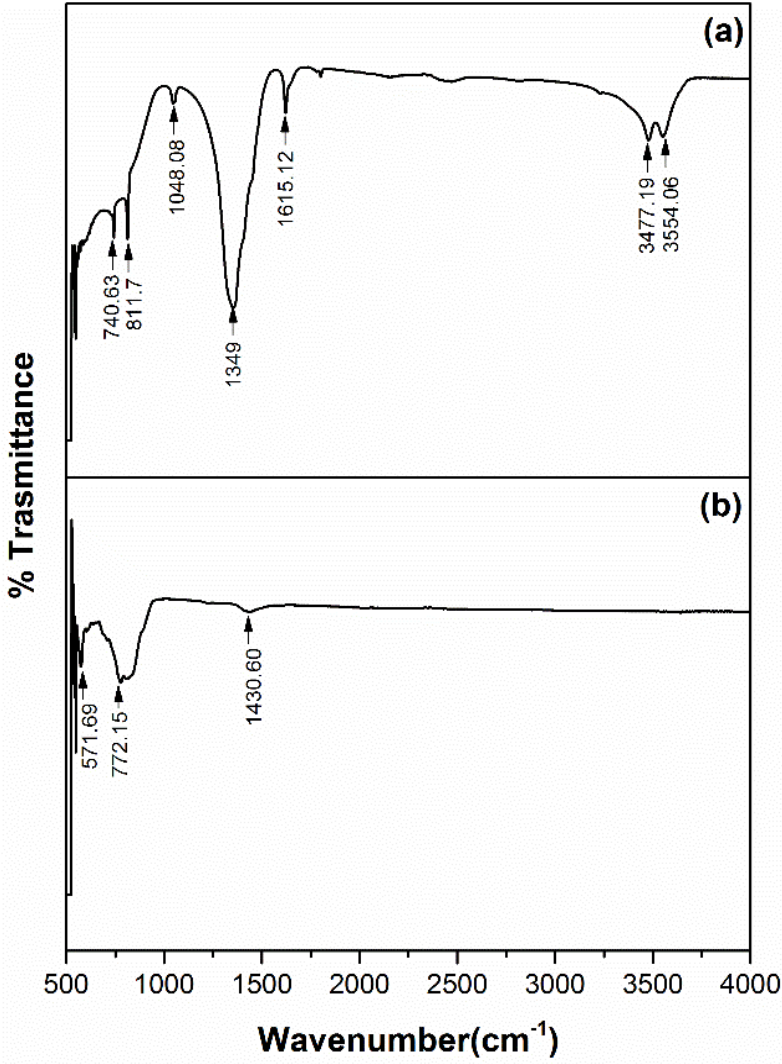

**FIGURE 8:** PL spectrum and (b) multipeak Gaussian fitting of C12A7 particles synthesized at 950 °C.

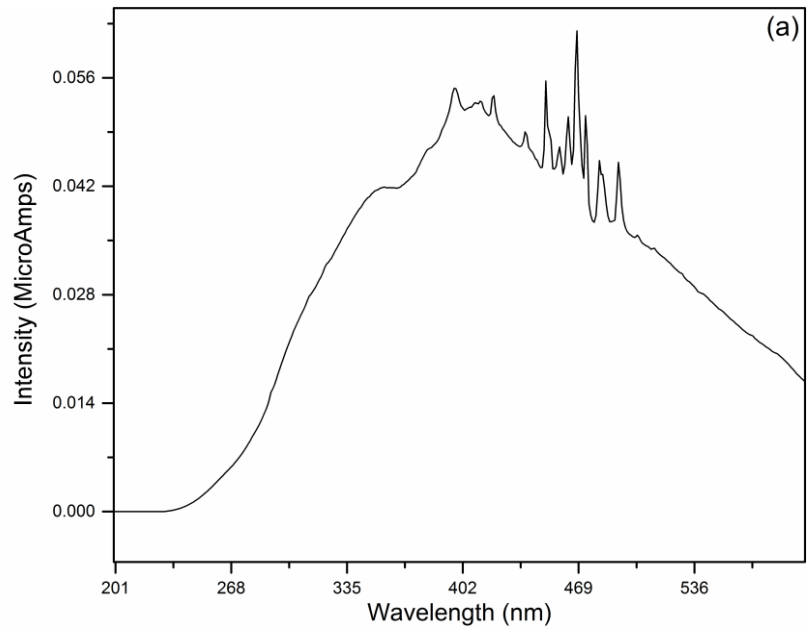

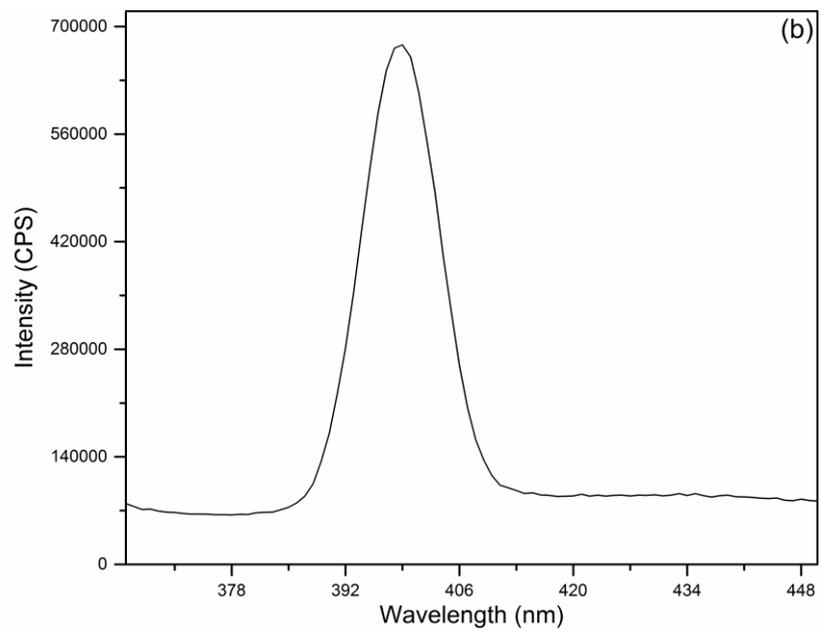

**Table 1**. Crystallite size and micro-strain of C12A7 particles.

| Sample | Synthesis temp. (°C) | Crystallite size (nm) | Micro strain |
|--------|----------------------|-----------------------|--------------|
| C12A7  | 510 °C               | 90                    | 0.00164      |
| C12A7  | 950 °C               | 96                    | 0.00153      |